\documentclass[fleqn,usenatbib]{mnras}

\usepackage{newtxtext,newtxmath}

\usepackage{graphicx}	
\usepackage{amsmath}	
\usepackage[T1]{fontenc}
\usepackage{fontawesome}
\usepackage{color}
\usepackage{amsmath}
\usepackage{natbib}
\usepackage{ctable}
\usepackage{bm}
\usepackage[normalem]{ulem} 
\usepackage{xspace}
\usepackage{paralist}
\usepackage{fontawesome}
\usepackage[caption=false]{subfig}

\let\oldAA\AA
\renewcommand{\AA}{\text{\normalfont\oldAA}}

\newcommand{\bitem}{\begin{itemize}}
\newcommand{\eitem}{\end{itemize}}
\newcommand{\beq}{\begin{equation}}
\newcommand{\eeq}{\end{equation}}

\definecolor{orange}{rgb}{1,0.5,0}

\title[\texttt{{CHARM}}]{\texttt{{CHARM}}: Creating Halos with Auto-Regressive Multi-stage networks}
\author[Pandey et al.]{Shivam Pandey,$^{1}$ 
Chirag Modi,$^{2,3}$
Benjamin D. Wandelt,$^{3,4}$
Deaglan J. Bartlett,$^{4}$
Adrian E.~Bayer,$^{5,3}$
\newauthor
Greg L. Bryan,$^{1}$
Matthew Ho,$^{1,4}$
Guilhem Lavaux,$^{4}$
T. Lucas Makinen,$^{6}$
Francisco Villaescusa-Navarro$^{3,5}$
\\$^{1}$Columbia Astrophysics Laboratory, Columbia University, 550 West 120th Street, New York, NY 10027, USA
\\$^{2}$Center for Computational Mathematics, Flatiron Institute, 162 5th Avenue, New York, NY 10010, USA
\\$^{3}$Center for Computational Astrophysics, Flatiron Institute, 162 5th Avenue, New York, NY 10010, USA
\\$^{4}$CNRS \& Sorbonne Universit\'{e}, Institut d’Astrophysique de Paris (IAP), UMR 7095, 98 bis bd Arago, F-75014 Paris, France
\\$^{5}$Department of Astrophysical Sciences, Princeton University, 4 Ivy Lane, Princeton, NJ 08544 USA
\\$^{6}$Imperial Centre for Inference and Cosmology (ICIC), Imperial College London, Prince Consort Road, London SW7 2AZ, United Kingdom
}

\date{Accepted XXX. Received YYY; in original form ZZZ}

\pubyear{2015}

\begin{document}
\label{firstpage}
\pagerange{\pageref{firstpage}--\pageref{lastpage}}
\maketitle

\begin{abstract}
To maximize the amount of information extracted from cosmological datasets, simulations that accurately represent these observations are necessary. However, traditional simulations that evolve particles under gravity by estimating particle-particle interactions ($N$-body simulations) are computationally expensive and prohibitive to scale to the large volumes and resolutions necessary for the upcoming datasets.
Moreover, modeling the distribution of galaxies typically involves identifying virialized dark matter halos, which is also a time- and memory-consuming process for large $N$-body simulations, further exacerbating the computational cost. In this study, we introduce \texttt{CHARM}, a novel method for creating mock halo catalogs by matching the spatial, mass, and velocity statistics of halos directly from the large-scale distribution of the dark matter density field. We develop multi-stage neural spline flow-based networks to learn this mapping at redshift $z=0.5$ directly with computationally cheaper low-resolution particle mesh simulations instead of relying on the high-resolution $N$-body simulations. We show that the mock halo catalogs and painted galaxy catalogs have the same statistical properties as obtained from $N$-body simulations in both real space and redshift space. Finally, we use these mock catalogs for cosmological inference using redshift-space galaxy power spectrum, bispectrum, and wavelet-based statistics using simulation-based inference, performing the first inference with accelerated forward model simulations and finding unbiased cosmological constraints with well-calibrated posteriors. The code was developed as part of the Simons Collaboration on Learning the Universe  and is publicly available at \url{https://github.com/shivampcosmo/CHARM}.
\end{abstract}





\section{Introduction}

The standard model of cosmology describes the evolution of the Universe using a set of free cosmological parameters. Constraining these parameters with observations of the Universe is one of the primary goals of cosmological studies \citep{bernardeau2002, alam2017, PlanckCollaboration:2020:A&A:, Albrecht:2006:arXiv:}. Over approximately 13.7 billion years of evolution, the hierarchical structure formation process transforms an initially Gaussian distribution of matter into a highly non-Gaussian field comprising halos, voids, and filaments. The observed galaxies occupy the collapsed and bound structures of dark matter called halos. The cosmological parameters can be constrained by analyzing the statistical distribution of the observed galaxies and comparing them to predictions from theoretical models or simulations. Traditional techniques limit this comparison to simple two-point summary statistics, such as the power spectrum at large scales, as theoretical models break down for higher-order statistics and non-linear small scales \cite{Nishimichi:2020:PhRvD:, philcox2022, damico2022, chen2022}. Since the evolved matter distribution is non-Gaussian, higher-order statistics, as well as small-scale two-point correlations, carry a significant amount of complementary information about the cosmological parameters \citep{simbigletter, massara2020, banerjee2021, Bayer:2021:ApJ:, eickenberg2022, valogiannis2022, naidoo2022, Makinen_2022}.

To extract this information, we need to rely on accurate $N$-body simulations as forward models and employ simulation-based inference (SBI) techniques \citep{alsing2019, Beyond-2ptCollaboration:2024:arXiv:,simbigletter, simbig_mock, Ho:2024:OJAp:}. SBI involves using computational forward models to simulate the data for a set of cosmological parameters, measuring the statistics of interest, and comparing them with the observed data using machine learning techniques to constrain the parameters (see \cite{cranmer2020} for a review). For galaxy clustering surveys, these forward models involve evolving the dark matter particles under gravity, identifying the dark matter halos, and then populating these halos with galaxies. However, simulating the high-mass halos (which are very rare) in various environmental conditions requires a large simulation box, while simulating the lower halo masses demands high resolution. Particularly, to reliably analyze the current generation of galaxy surveys, the number of particles and the volume of the simulation required are so large that the computational cost of running these simulations at a grid of cosmological parameters is prohibitive. To put things in context, analyzing the last generation of cosmological surveys, which ended a decade ago, with this approach would require running at least 2000 simulations with $2.7\times 10^{10}$ particles, taking more than 270 million CPU hours for running the simulations alone \citep{villaescusa-navarro2020}. Furthermore, finding the halos accurately, which requires processing phase space information of particles \citep{behroozi2013a} in these $N$-body simulations, also adds to the computational cost.

However, physically, we expect the number, mass, and velocity distributions of halos to depend on the large-scale matter distribution (see \cite{desjacques2016} for a review). For example, the overdense regions of the Universe will have more matter to collapse and will be able to form more numerous and massive halos. Therefore, accurately learning this relationship and generating fast approximations of the dark matter distribution on large scales can accelerate mock halo catalog generation, and ultimately generate observed data with end-to-end simulations. This motivates us to use deep learning techniques to learn these highly non-linear and non-local \citep{Bartlett:2024:arXiv:} relationships between the dark matter and halo distributions.

To accelerate the simulations on large scales, particle mesh (PM) approximations can be used \citep[e.g.,][]{Tassev2013, Feng2016}. These approximations estimate the gravitational forces by interpolating CDM (cold dark matter) particles on a uniform grid, enabling the use of techniques such as fast Fourier transforms to solve the equations of motion. Due to this grid interpolation, they lose information on scales smaller than the grid resolution, resulting in poor halo catalogs \citep{Wu:2024:MNRAS:, Doeser:2023:arXiv:}, particularly for low-mass halos. However, on scales larger than the grid size, they accurately capture the matter distribution. Since these PM simulations are orders of magnitude faster than $N$-body simulations, our goal is to learn the relationship between $N$-body halos, and matter density fields obtained from paired (ran with same initial conditions) low-resolution PM simulations.

In this work, we introduce \texttt{CHARM}: a generative model for creating halo catalogs using multi-stage neural spline flows to transform the low-resolution PM simulations to discrete mock catalogs expected from a high-resolution $N$-body simulation. It consists of four stages after extracting features from the surrounding dark matter density in PM simulations at any location: $(i)$ learn the number of halos expected, $(ii)$ learn the mass of the heaviest halo, $(iii)$ auto-regressively learn lower halo masses, and $(iv)$ auto-regressively learn the 3D halo velocities. Developing a methodology like this which is accurate, is fast enough to be used for cosmological inference, and is scalable to larger volumes is crucial for using simulation-based inference techniques to analyze large volume data and maximize the information gained from current and future galaxy surveys, a key goal of, for example, the Learning the Universe collaboration\footnote{\url{https://learning-the-universe.org}}. We will test the performance of our model across all of these three axes -- accuracy, inference, and scalability for the analysis of galaxy survey data.

In recent years, there have been other studies with related goals, but they provide different solutions compared to what is desired here. In \cite{charnock2020} and \cite{Ding:2024:arXiv:}, a similar mapping is learned using physically motivated networks with a significantly reduced number of free parameters, but they assume an explicit form of the likelihood for halo occupation, which breaks down for high-mass halos and small scales, which are of interest in this study. Moreover, these models have been trained on a fixed cosmology simulation in real space. In contrast, we aim to obtain a forward model that generalizes to different cosmologies and to redshift space. There have also been attempts, as in \cite{Modi_2018} and \cite{zhang2019dark}, that do not impose a likelihood form. However, the methodology of \cite{Modi_2018} only works for continuous biased scalar fields like total halo mass, whereas here we aim to obtain discrete halo catalogs. The methodology of \cite{zhang2019dark} is designed to work only with dark matter density obtained from high-resolution $N$-body simulations, thus requiring large computational resources.

In \cite{jamieson2022, Jamieson:2024:arXiv:}, displacement corrections to PM simulations were provided to make them resemble their $N$-body counterparts. However, these corrections, applied at the particle level, only improve the statistics of relatively high mass halos ($M > 10^{14} M_{\odot}/h$), are calibrated on relatively low-resolution simulations, and are expensive to train. Note that the corrections can be improved if additional force evaluations are included \citep{Bartlett_2024_COCA}.
In \cite{Wu:2024:MNRAS:}, the authors present methods to correct PM simulations, leading to a faithful reproduction of halos even on small scales. However, they require high-resolution PM runs, whereas here we focus on obtaining $N$-body-like halo catalogs while using low-resolution PM simulations that can be scaled to large volumes. Nevertheless, these corrections could be used to augment the PM simulations used here and improve the accuracy of the model in the future.

The paper is organized as follows: in \S~\ref{sec:data} we describe the simulation dataset used, in \S~\ref{sec:methods} we describe the architecture of our model, in  \S~\ref{sec:stats_inference} we describe the setup for evaluating the performance of the model, in \S~\ref{sec:results} we show the results and finally conclude in \S~\ref{sec:discussion}. 

\section{Dataset}
\label{sec:data}

\paragraph*{Simulations:}
We use the public simulation suite from the $N$-body Quijote project \citep{villaescusa-navarro2020}, which simulates a volume of $(1000{\rm Mpc}/h)^3$. These simulations cover a wide range of cosmologies and have enough volume and resolution to provide reliable dark matter halo catalogs with masses above $\sim 5\times 10^{12} M_{\odot}/h$. Therefore, these simulations provide a good suite to build a reliable model of how the distribution of halos relates to the surrounding dark matter field, which can then be applied to significantly larger volume simulations. In this work, we process the high-resolution Latin hypercube (LH) and fiducial cosmology suite of simulations, varying five cosmological parameters: matter density $\Omega_{\rm m}$, baryon density $\Omega_{\rm b}$, matter clustering strength $\sigma_8$, primordial power spectrum tilt $n_s$, and the expansion rate of the Universe $h$. These five cosmological parameters are varied with a wide uniform prior of: $\Omega_{\rm m} \in \mathcal{U}[0.1, 0.5]$, $\sigma_8 \in \mathcal{U}[0.6, 1.0]$, $\Omega_{\rm b} \in \mathcal{U}[0.03, 0.07]$, $n_s \in \mathcal{U}[0.8, 1.2]$, and $h \in \mathcal{U}[0.5, 0.9]$. In this study, we use only the density fields and halo catalogs at $z=0.5$. We use sub-volumes from 1800 LH simulations and one full fiducial cosmology simulation for training while reserving the remaining 200 LH simulations for testing. Each $N$-body simulation evolves an initial Gaussian distribution of $1024^3$ cold dark matter particles to the present time and takes approximately 5000 CPU hours to complete.

\paragraph*{Input dark matter fields:}
For the approximate PM simulations, which form our input, we run paired simulations (i.e., matching Gaussian initial conditions and cosmology to the Quijote suite) using the \texttt{\texttt{FastPM}} algorithm \citep{Feng2016}. Here, we evolve only $384^3$ particles over the same volume, and each simulation takes only 5 CPU hours\footnote{GPU implementations of these algorithms can further increase computational efficiency \citep{modi2021, Li2022}}. We compute the matter density fields ($\rho_{\rm m}$) and 3D matter velocity fields ($[{v}^x_{\rm m}, {v}^y_{\rm m}, {v}^z_{\rm m}]$) from the PM simulations on a regular grid with $128^3$ voxels using cloud-in-cell interpolation. Note that each voxel has a physical size of approximately 7.8 Mpc$/h$. It is also important to note that, since the size of a typical halo is less than 1 Mpc/$h$, each voxel can host multiple halos. We concatenate the density and velocity fields, which form the input to our model.

\paragraph*{Target halo catalog:}
From the $N$-body simulations, we use the halo catalog obtained by applying the \texttt{Rockstar} halo finder to the particle distribution. The algorithm uses the 6D phase space information of the particles to identify particles in collapsed structures. We use halos with masses $M_{\rm 200c} \geq 5\times 10^{12} M_{\odot}/h$ in the catalog, which is the resolution limit of the Quijote simulations. Note that the spherical overdensity mass, $M_{\rm 200c}$, is the mass within $r_{\rm 200c}$ of the halo which is defined such that the average enclosed density within a sphere of radius $r_{\rm 200c}$ is equal to $200$ times the critical density of the Universe, $\rho_c(z)$. The low-resolution PM simulations used here do not account for small-scale particle-particle interactions, resulting in biased small-scale matter velocity and density fields. As \texttt{Rockstar} relies on an accurate small-scale velocity distribution of matter particles, it heavily underestimates the distribution of low-mass halos \citep{Wu:2024:MNRAS:, Doeser:2023:arXiv:}. However, the spatial distribution and velocity of these halos are correlated with the large-scale fields (which are correctly captured by the PM simulations, e.g., see \cite{Bayer:2023:JCAP:}), and we aim to learn this mapping using \texttt{CHARM}.

We voxelize our target halo distribution on the same $128^3$ grid using the nearest-grid point (NGP) mass assignment scheme. Within this scheme, for each voxel $i$, we count the number of halos inside it ($N^i_{\rm tot}$). If it is non-zero, we also store the halo masses in decreasing order ($[M^i_1, M^i_2, \ldots, M^i_{N^i_{\rm tot}}]$, where $M^i_1 > M^i_2 > \ldots > M^i_{N^i_{\rm tot}}$). In our training set, we have at most approximately 12 halos in any voxel, so we fix $N_{\rm max} = 12$, and $N^i_{\rm tot} \leq N_{\rm max}$ for all $i$.

While the PM simulations underestimate the small-scale velocity distribution, they capture the large-scale coherent velocity fields correctly. Therefore, we estimate the 3D PM velocity for each halo by linearly interpolating the 3D matter velocity field from PM simulations ($\vec{v}_{\rm {\texttt{FastPM}}}$) and aim to estimate the difference, $\Delta \vec{v} = \vec{v}_{\rm Quijote} - \vec{v}_{\rm {\texttt{FastPM}}}$, for each halo in the voxel. We order the halos in the same order as for masses, going from heaviest to least heavy.

\paragraph*{Batching:}
We divide the 3D simulation boxes, each of size $128^3$, into sub-boxes of size $16^3$, resulting in 512 sub-boxes from each simulation. Each sub-box has a physical size of 125 Mpc/$h$. We select the corresponding sub-boxes from the PM set to create the input dark matter density field, paired with the $N$-body halos for training. To facilitate feature extraction, we pad the input density field from the PM sub-boxes so that the output after convolutions preserves the size of the sub-box ($16^3$). Therefore, we add a padding of four voxels on each side from the original periodic simulations.

For training, we use one full simulation (512 sub-boxes) at the fiducial cosmology while using only 16 randomly selected sub-boxes for 2,000 simulations with varying cosmologies in the LH grid (using only about $\sim 3\%$ of the available volume). This is done primarily due to the memory limitations of GPUs and to accelerate the training process. With one full simulation at a fiducial cosmology, the network learns the dependence of halo properties on the environment at a fixed cosmology, while with the limited data from the LH set, the network learns the variation of properties with cosmological parameters.

\begin{figure*}
  \centering
 \includegraphics[width=1.\textwidth]{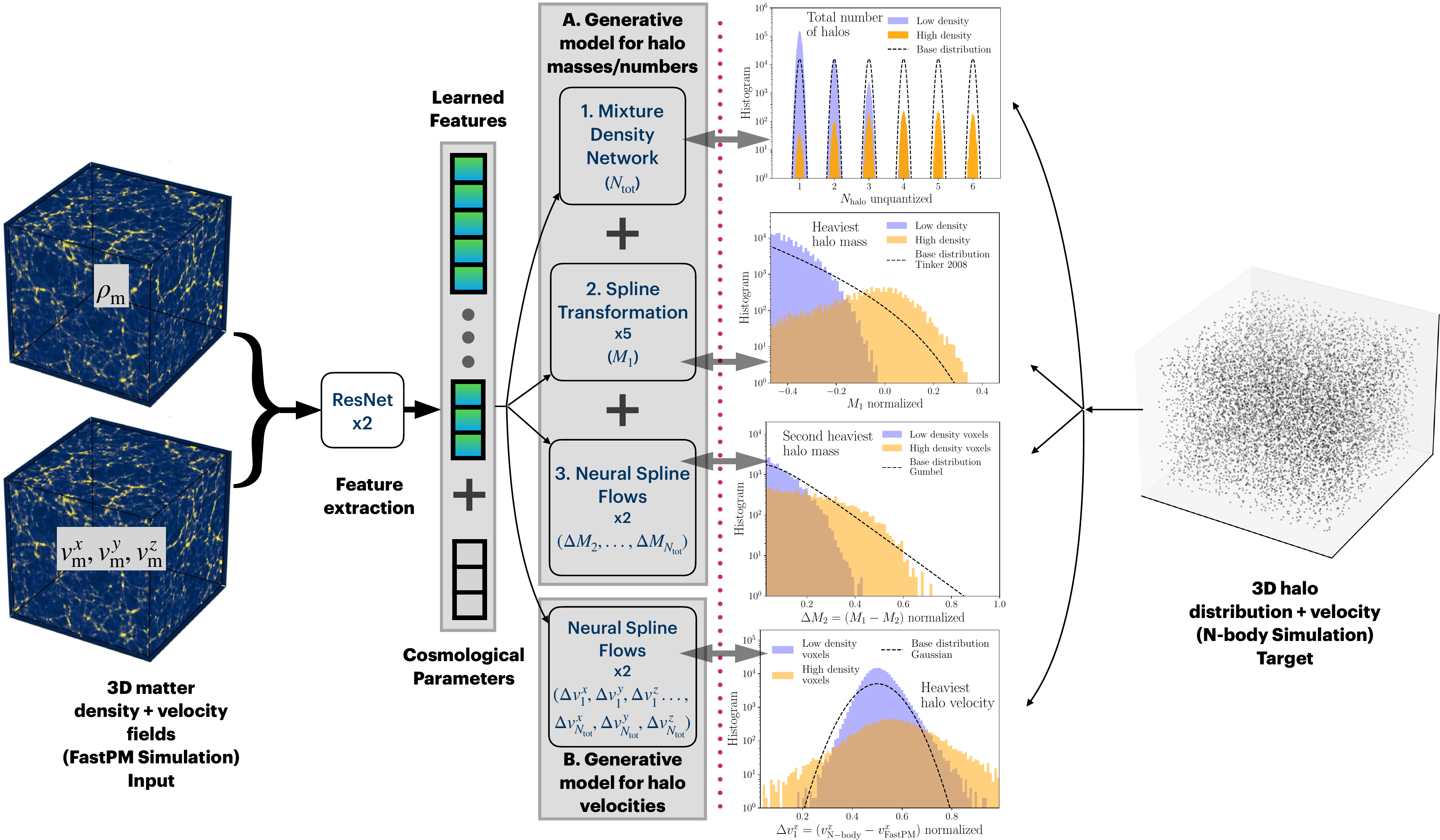}  
  \caption{Visualization of the data products and network architecture used in this study. On the left side of the dotted line, we show the matter density and velocity fields from the PM simulation that serve as the input to the ResNet layers to extract features. These features are then used to predict the halo distribution and velocities in four parts: the total number of halos is modeled using a mixture density network, the heaviest halo mass is modeled using a stack of spline transformations, and lower halo masses as well as halo velocities are modeled using a stack of auto-regressive neural spline flows. On the right side, we display the target distribution of the halos from paired true $N$-body simulation. Additionally, we present histograms of the four quantities in both low and high-density voxels, illustrating their dependence on the dark matter distribution.}
  \label{fig:network}
\end{figure*}

\section{Methodology} 
\label{sec:methods}

The task of this paper is to obtain a discrete mock halo catalog when provided with approximate dark matter over-density and velocity fields from PM simulations. Halo formation is a complex non-linear process that depends on the 3D matter distribution on large scales. For example, there is a higher probability of forming a heavy halo at the intersection of large dark matter filaments compared to in a void region. Therefore, to extract the features of dark matter density and velocity fields that correlate with the halo distribution, we stack two 3D residual network (ResNet) layers \citep{he2015deep}. We input the dark matter density and 3D velocity fields as described in Section.~\ref{sec:data} as four channels and extract 20 features from ResNet layers. These features, extracted from a physical region of approximately 70 Mpc/$h$, are used as conditioning for a multi-level generative model for the halo distribution, as described below. Note that we also append the values of the five cosmological parameters for each simulation sub-box to these 20 features.

To create a mock halo catalog and its masses, we need to estimate a discrete distribution of halos conditioned on a feature vector for each voxel. To achieve this, we split the problem into three parts. For each voxel $i$, we first predict the total number of halos ($N^i_{\rm tot}$), which provides a mask as well as an occupation number to train the mass distribution. Then, for the voxels that have a non-zero number of halos, we predict the mass of the heaviest halo ($M^i_1$). Afterward, for voxels that have more than one halo, we train the prediction for the masses of lighter halos ($M^i_2, \ldots , M^i_{N^i_{\rm tot}}$) in an auto-regressive fashion. This means that, as dictated by the physics of structure formation, we always condition the probability of lighter halo masses on the masses of all the heavier halos in the same voxel. Our final loss function is a sum of the losses from these three steps and the details of each step are as follows (also see Fig.~\ref{fig:network} for a summary of this inference pipeline):

\begin{enumerate}
    \item We model the probability distribution of the total number of halos as a mixture of $N_{\rm max}$ Gaussians. We take the discrete distribution of the number of halos in each voxel of the training simulations and make it continuous by adding a small Gaussian noise with a known variance ($\sigma^2 = 0.0025$). With input from the learned features of the ResNet, we predict the probability of each Gaussian using a fully connected neural network (FCNN). In this case, the loss is modeled as the forward Kullback–Leibler (KL) divergence \citep{Kullback_Leibler:1951} between the modeled Gaussian mixture with predicted probabilities (with fixed mean and variance) and the true distribution.
    \item To model the mass of the heaviest halo, we first transform a base distribution using a stack of five spline transformations \citep{durkan2019neural} with 8 knots. 
    The base distribution here is not the standard Gaussian, as is often used with normalizing flows, but the probability distribution function estimated from the unconditional halo mass function, as described in \cite{Tinker_2008} (see the upper-central black-dashed histogram in Fig.~\ref{fig:network}). We found that using this physically motivated base distribution was crucial for obtaining accurate predictions for the heaviest halo.
    The parameters of the transformations are learned using a separate FCNN. The loss is then calculated as the KL divergence between a known base distribution and this transformed distribution. 
    \item We learn the distribution of lower halo masses using an auto-regressive neural spline flow \cite{durkan2019neural}. We condition the transformation on the masses of all the heavier halos. Additionally, to ensure a decreasing order of halo masses, for the $j$-th halo in the voxel, we learn the mass difference $M_{j-1} - M_{j}$ and ensure that this difference is positive. Here, we stack two such spline flows, and their parameters are once again learned using an FCNN. As having more halos in the same voxel becomes a rarer phenomenon, we model the base distribution as proportional to the Gumbel distribution \cite{gumbel}, which provides a good initial estimate of extreme value statistics (see the bottom-central histogram in Fig.~\ref{fig:network}). 
\end{enumerate}

Finally, we train an independent auto-regressive network to obtain the velocities of the halos conditioned on the features extracted from the PM simulation. As described in \S~\ref{sec:data}, we use the difference between the true halo velocity and its interpolated value estimated from the PM simulation. The PM simulation velocity fields, smoothed at $\sim 8\, {\rm Mpc}/h$ resolution already capture the large-scale coherent velocity component of the halos. Therefore, with this network, we learn the remaining small-scale component of the velocities (related to the fingers-of-god effect, \cite{Jackson:1972:MNRAS:}). We start with a Gaussian distribution as our base distribution and use a stack of two neural spline flows to learn the 3D velocity difference of halos at each voxel. 

Once trained, this model can then be used to obtain a catalog of mock halo positions in real space ($\vec{x}_h$), their masses, and their velocities ($\vec{v}_h$) on test simulations by inputting the corresponding PM density and velocity fields. To bring the model closer to the observations, we need to place the objects in redshift space. With the knowledge of $\vec{x}_{\rm h}$ and $\vec{v}_{\rm h}$, we can calculate the position of halos in redshift space ($\vec{s}_{\rm h}$) with $\vec{s}_{\rm h} = \vec{x}_{\rm h} + (\eta_{\rm l.o.s}) \cdot (\vec{v}_{\rm h} \times (1 + z)/H(z))$, where $\eta_{\rm l.o.s}$ is the line of sight direction (which we assume to be parallel to the $x-$axis of the box in this study) and $H(z)$ is the Hubble constant at redshift $z$.

\section{Diagnostics and inference setup} 
\label{sec:stats_inference}

\subsection{Statistical summaries}
\label{sec:stats_methods}
Once the model is trained, we obtain mock halo and galaxy catalogs on test LH simulations and compare their performance by calculating various statistics of interest for cosmological studies. In addition to comparing the environment-dependent histograms of mock and true halo catalogs, we also compare the higher-order statistical properties. We calculate three different statistics that are sensitive to two-point, three-point and higher-order correlations between the galaxies.

\paragraph*{Power spectrum:} We calculate the real space 3D power spectrum on 16 scales between $k_{\rm min} = 0.01$ and $k_{\rm max} = 0.32$ and use the nearest-grid point estimator to calculate the density fields. We use the public \texttt{Pylians} code to estimate the power spectra.\footnote{\url{https://pylians3.readthedocs.io/en/master/}} In redshift space, we calculate and compare the power spectra multipoles $P_{\ell}(k)$ for $\ell \in [0,2,4]$. 

\paragraph*{Bispectrum:} We calculate the bispectrum for isosceles triangles for three different scales, $k \in [0.08, 0.16, 0.32] h/{\rm Mpc}$ and eight different values of angle between the equal edges, ranging between 0.1~radians to 3.04~radians. These are also calculated using the \texttt{Pylians} code. In redshift space, we only calculate and compare the monopole of bispectrum $B_{\ell = 0}(k)$.  

\paragraph*{Wavelets:}
We follow \cite{Regaldo-SaintBlancard:2024:PhRvD:} for calculating the wavelet scattering statistics using their public code.\footnote{\url{https://github.com/bregaldo/galactic\_wavelets/tree/main}} Given the 3D density field ($\rho$) constructed from the mock galaxy or halo catalogs, we calculate three different wavelet coefficients by convolving the halo field with various wavelets and evaluating the moments:
\begin{align}
    S_0 &= \langle |\rho|^p \rangle \\
    S_1 &= \langle |\psi_{\lambda_0} \ast \rho|^p \rangle \\
    S_2 &= \langle ||\psi_{\lambda_1} \ast \rho| \ast \psi_{\lambda_2}|^p \rangle,
\end{align}
where, $\psi_{\lambda_j}$ is the wavelet, $\lambda_j$ is its index, and $p$ is the exponent. The wavelets are constructed from a directional mother wavelet which is defined in Fourier space. Its form is localized in real space and has a cut-off scale ($k_c$) in Fourier space. We set $k_c = 4\pi k_{\rm max}/3$, where we again use $k_{\rm max} = 0.32 \, h/{\rm Mpc}$. We set the number of octaves (doublings of scales) to $J=3$ and set $Q=4$ which controls the number of scales per octave. We evaluate the wavelet coefficients for three different values of exponent $p \in [1/2, 1, 2]$.

\subsection{Galaxies}
\label{sec:galaxies}

Instead of dark matter halos, we typically observe galaxies and so we need to predict the statistics of galaxies from a given halo catalog. There are several ways to establish the halo-galaxy connection ranging from simple empirical fitting functions \citep{Conroy:2006:ApJ:, Behroozi:2010:ApJ:, Moster:2010:ApJ:, zheng2007,Yang:2009:ApJ:,Moster:2018:MNRAS:} to more physical models (see \cite{Somerville:2015:ARA&A:} for a review) that describe the galaxy formation through cosmic times using hydrodynamical simulations \citep[e.g.,][]{Vogelsberger:2014:MNRAS:, villaescusa-navarro2022} or semi-analytic models \citep[e.g.,][]{Somerville:1999:MNRAS:}. These methods have a wide range of applicability and complexity and in this study where we aim to target spectroscopic red galaxy catalogs, such as the CMASS galaxy sample of the Sloan Digital Sky Survey (SDSS-CMASS), we choose the empirical halo occupation distribution (HOD) approach. This approach has been successfully applied in several analyses for both standard two-point power spectra \cite[e.g.,][]{reid2014} and higher order correlations \citep[e.g.,][]{Hahn:2024:PhRvD:, Regaldo-SaintBlancard:2024:PhRvD:}.
In this first study to use an accelerated forward model for halos, we use a simple five-parameter HOD model (\cite{zheng2007}) which assumes that the occupation probability of galaxies in any halo depends only on its mass. We defer the study with a more realistic halo-galaxy connection to a future investigation.

The \cite{zheng2007} HOD model parameterizes the number of central ($N_c$) and satellite ($N_s$) galaxies as a function of halo mass with five free parameters: ${\log(M_{\rm min}), \sigma_{\log M}, \log(M_0), \log(M_1), \alpha_{\rm sat} }$. We sample 10 random values for the HOD parameters for each LH simulation, resulting in 20,000 mock galaxy catalogs. The parameters $\sigma_{\log M}$ and $\alpha_{\rm sat}$ are sampled with wide uniform priors: $\sigma_{\log M} \in \mathcal{U}[0.3, 0.5]$ and $\alpha_{\rm sat} \in \mathcal{U}[0.3, 0.5]$. The other three HOD parameters are constrained to vary with the cosmology ($\vec{\theta}_{\rm cosmo}$) of each LH simulation: $\log(M_{\rm min}) \in \mathcal{U}[\log(M^{\vec{\theta}_{\rm cosmo}}_{\rm min}) \pm 0.15]$, $\log(M_{0}) \in \mathcal{U}[\log(M^{\vec{\theta}_{\rm cosmo}}_{0}) \pm 0.2]$, and $\log(M_{1}) \in \mathcal{U}[\log(M^{\vec{\theta}_{\rm cosmo}}_{1}) \pm 0.3]$. The cosmology-dependent central values of the priors for these parameters, ($\log(M^{\vec{\theta}_{\rm cosmo}}_{\rm min}), \log(M^{\vec{\theta}_{\rm cosmo}}_{0}), \log(M^{\vec{\theta}_{\rm cosmo}}_{1})$), are fixed such that the number density of the mock galaxy catalog approximately matches the CMASS sample comoving galaxy density, $\bar{n}_g \sim 3\times 10^{-4} ({\rm Mpc}/h)^{-3}$, and the satellite fraction $f_{\rm sat} \sim 0.2$ \citep{reid2014}.

Assuming $\bar{n}_g$ and $f_{\rm sat}$, we estimate the expected number of central ($\bar{N}_c$) and satellite ($\bar{N}_s$) galaxies. Then, we calculate the minimum halo mass ($\log M^{\vec{\theta}_{\rm cosmo}}_h$) that results in $\bar{N}_c$ halos, each hosting a central galaxy, and set $\log(M^{\vec{\theta}_{\rm cosmo}}_{\rm min}) = \log(M^{\vec{\theta}_{\rm cosmo}}_{0}) = \log M^{\vec{\theta}_{\rm cosmo}}_h$. Finally, assuming a fiducial value of $\alpha_{\rm sat} = 0.7$, we determine $\log(M^{\vec{\theta}_{\rm cosmo}}_{1})$ that results in $\bar{N}_s$ satellite galaxies.

We assume that the central galaxy, which is placed at the center, has the same velocity as the parent halo, whereas the satellite galaxies are distributed around the halo following the Navarro-Frenk-White profile \citep{navarro1997} and receive an additional velocity contribution due to their virial motion. We place the galaxies at the real-space positions of the halos, assign their velocities, and then move them to redshift space. We then measure the statistical summaries from this distribution as described below.

However, note that \texttt{CHARM} does not resolve the positions of the halos within the $8 \, {\rm Mpc}/h$ voxel, so it places them at the voxel centers. To account for this lack of knowledge of galaxy positions on smaller scales, we apply smoothing to the positions of the galaxies in redshift space by adding a random Gaussian scatter to their 3D positions with a standard deviation of $\sigma_{\rm G} = 8\,{\rm Mpc}/h$. We apply this same smoothing to the true galaxy catalogs in redshift space generated from the test Quijote simulations, against which we compare the performance of the \texttt{CHARM} model.

\begin{figure*}
  \centering
 \includegraphics[width=1.\textwidth]{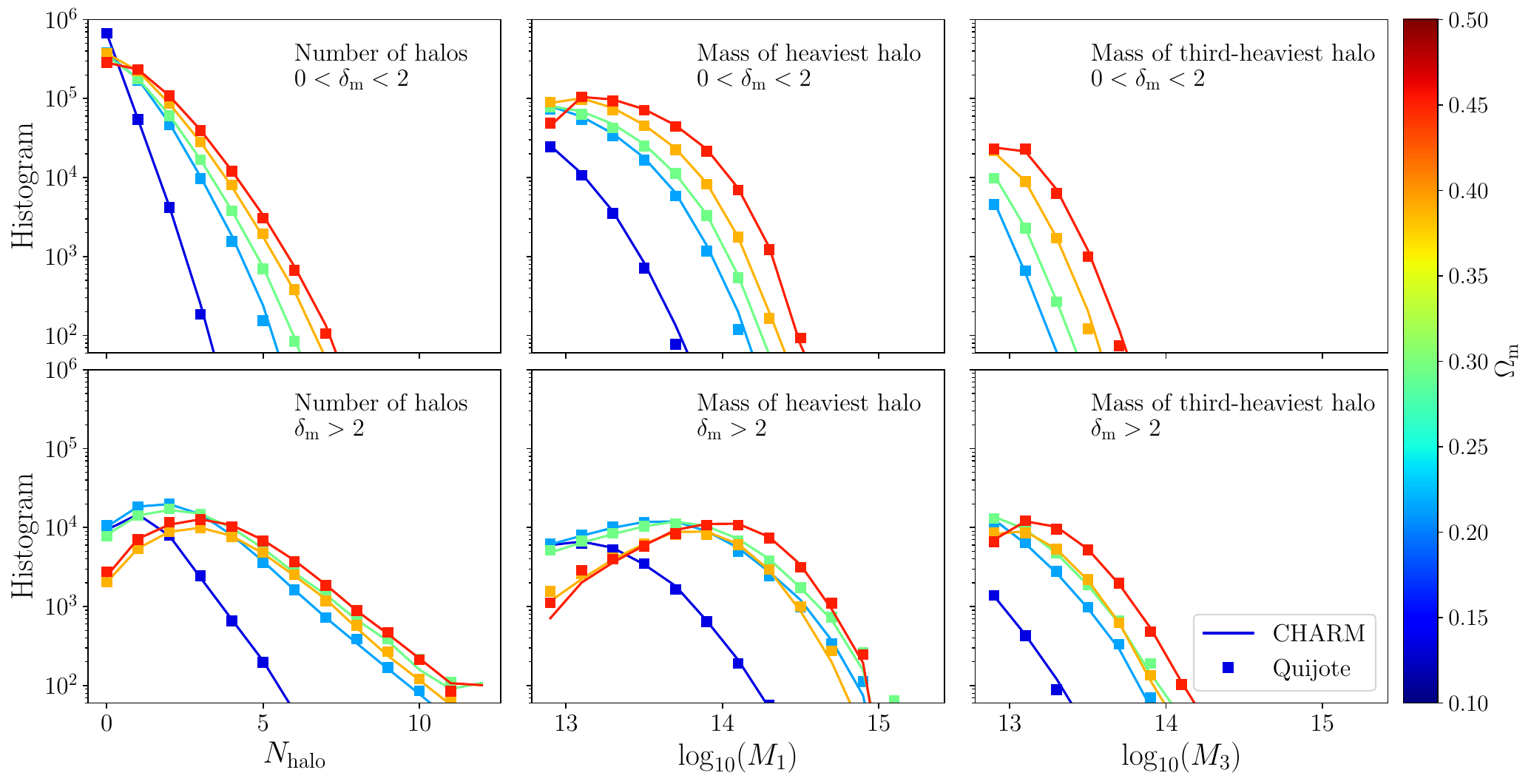}  
  \caption{\textbf{Local performance of the model with halos in real space:} Comparison of the one-point statistics between the true (square markers) and mock (solid lines) halo catalogs in test simulations with different cosmologies. The left, middle, and right columns compare the number of halos, the mass of the heaviest halo, and the mass of the third heaviest halo, respectively, in $128^3$ voxels for each cosmology. In the top row, we show the histogram for sub-selections of voxels with dark matter density in the range $0 < \delta_{\rm m} < 2$, while in the bottom row, we use high-density voxels with $\delta_{\rm m} > 2$. This test individually compares the performance of each of the three stages of the halo mass network (as described in \S~\ref{sec:methods}), demonstrating the high fidelity of the mock catalogs obtained using \texttt{CHARM}.}
  \label{fig:1pt}
\end{figure*}

\subsection{Inference setup}
\label{sec:inference}
We use neural posterior estimation (NPE) to obtain the posteriors on (cosmological and HOD) parameters of our forward model of the statistics of galaxies in redshift space. The NPE algorithm trains a conditional normalizing flow ($q_{\phi}(\vec{\theta}|\vec{x}_{\rm data})$), where $\vec{\theta}$ are the model parameters, $\vec{x}_{\rm data}$ is the data-vector of summary statistics compressed using an embedding network and $\phi$ are the hyper-parameters of the normalizing flow model and embedding networks. We use the standard masked auto-regressive architecture \citep{papamakarios2017masked} for normalizing flow which approximates the true posterior distribution ($p(\vec{\theta}, \vec{x})$) by approximating it with a sequence of auto-regressive affine transformations of a base distribution, with known probability density (here a standard Gaussian). The flow is trained by minimizing the Kullback-Leibler (KL) divergence:
\begin{align}
    \mathcal{L}(\phi) = D_{\rm KL}[p(\vec{\theta}, \vec{x})||q_{\phi}(\vec{\theta}|\vec{x}_{\rm data}) p(\vec{x}_{\rm data})],
\end{align}
which can be shown to be equivalent to maximizing the training score:
\begin{align}
    \mathcal{S}(\phi) = \sum_{j\in {\rm Train}} \log q_{\phi}(\vec{\theta}_j|\vec{x}_{\rm data, j}),
\end{align}
where the summation runs over all the training examples. We use the public \texttt{sbi} package to perform this inference \citep{tejero-cantero2020}.\footnote{\url{https://github.com/sbi-dev/sbi}}

To maximize the training score, we first train 256 networks with random values of hyperparameters $\phi$ of the network. The hyperparameters include the number of fully connected layers, hidden features and batch size to obtain parameters of affine transformations, as well as parameters of the embedding network compressing the input data-vector. We then select the five configurations with the best validation loss and generate an ensemble flow with these five trained networks by linearly combining them with uniform weights. We defer further optimization of this procedure, such as using a more optimal weighting scheme to combine the flows in an ensemble and using more expressive affine transformations such as neural spline flows, to a future study.

\begin{figure*}
  \centering
 \includegraphics[width=1.\textwidth]{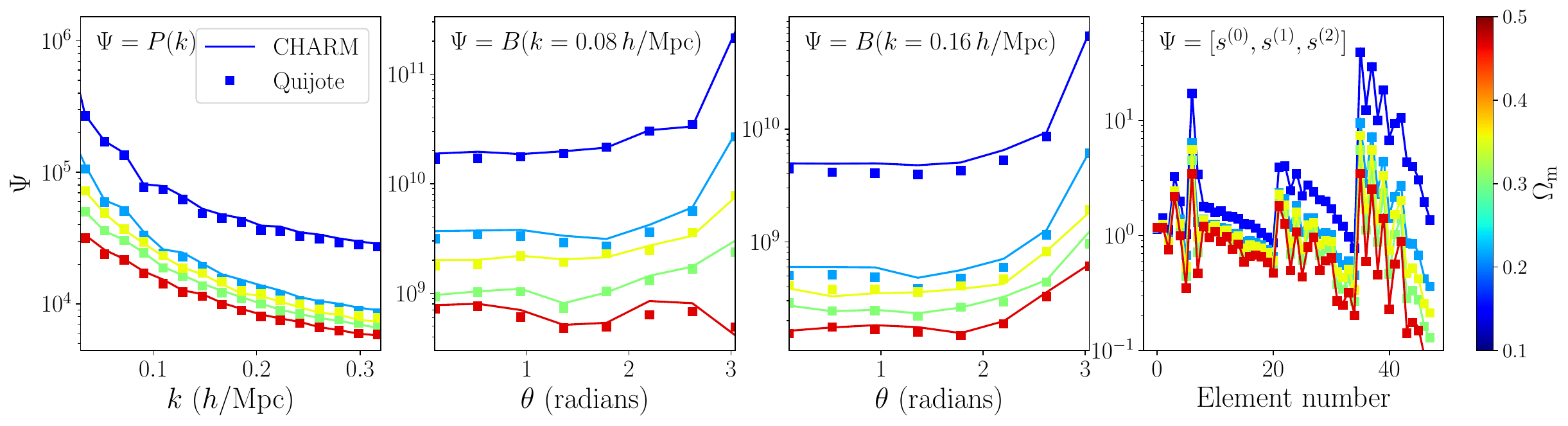}  
  \caption{\textbf{Global performance of the model with halos in real space:} Comparison of the mass-weighted power spectrum (left), bispectrum on large scales (middle left), bispectrum on small scales (middle right), and wavelet scattering transform (right) statistics between the true (square markers) and mock (solid lines) halo catalogs in real comoving space across five different test simulations, colored by the value of the $\Omega_{\rm m}$ cosmological parameter. When calculating the density field to compute these statistics, each halo is assigned a mass-dependent weight of $w = (M/M')^{\alpha}$, where we fix $M' = 10^{14} \, M_{\odot}/h$ and $\alpha = 0.7$ to mimic the effect of galaxies. }
  \label{fig:npt_real_halos}
\end{figure*}

\begin{figure*}
  \centering
 \includegraphics[width=1.\textwidth]{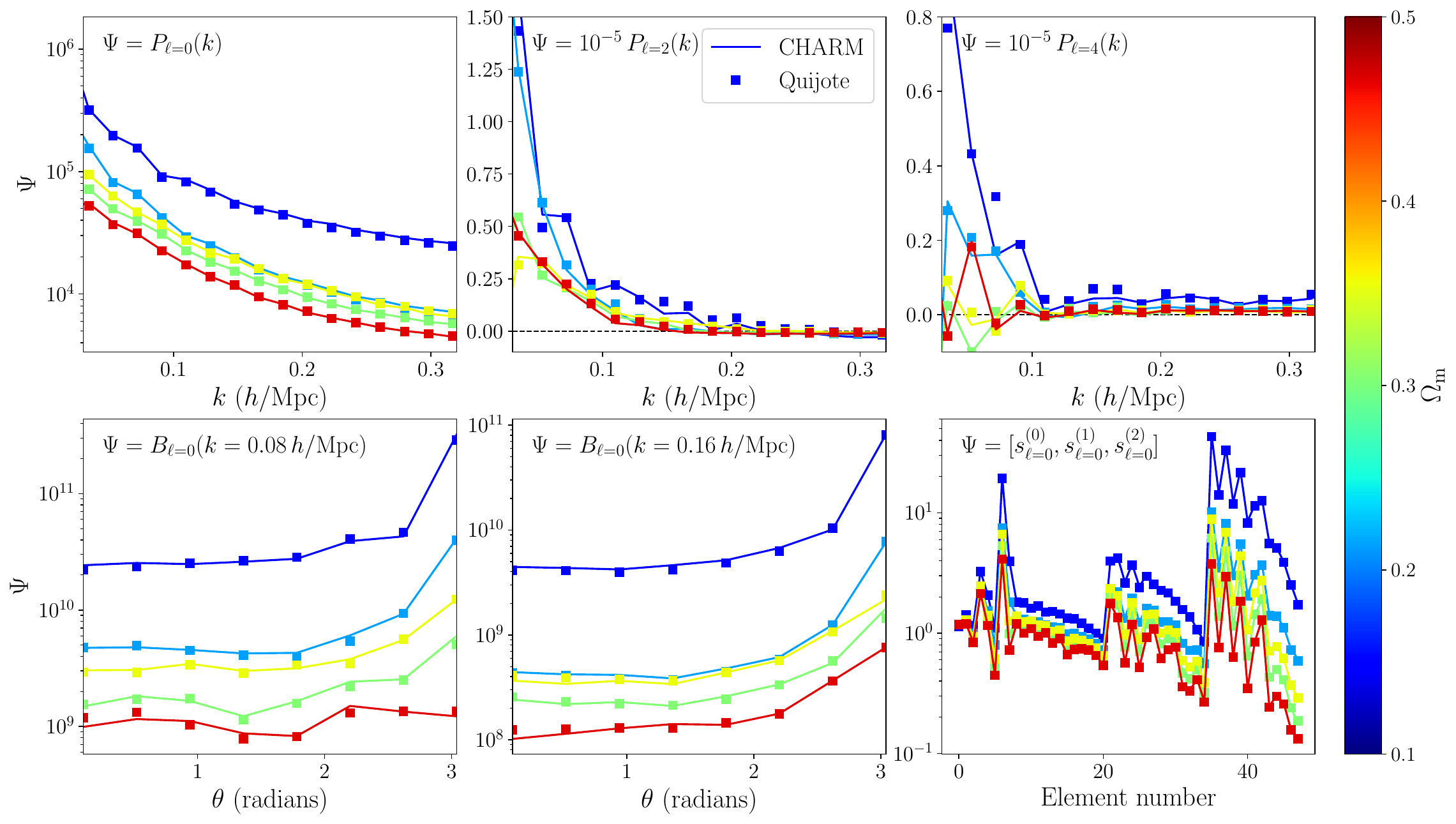}  
  \caption{\textbf{Global performance of the model with halos in redshift space:} Comparison of the power spectrum multipoles (top row), bispectrum monopole on large scales (bottom row, left column), bispectrum monopole on small scales (bottom row, center column), and wavelet scattering transform (bottom row, right column) statistics between the true (square markers) and mock (solid lines) halo catalogs in the test LH simulations, colored by the value of the $\Omega_{\rm m}$ cosmological parameter.}
  \label{fig:npt_rsd_halos}
\end{figure*}

\begin{figure*}
  \centering
 \includegraphics[width=1.\textwidth]{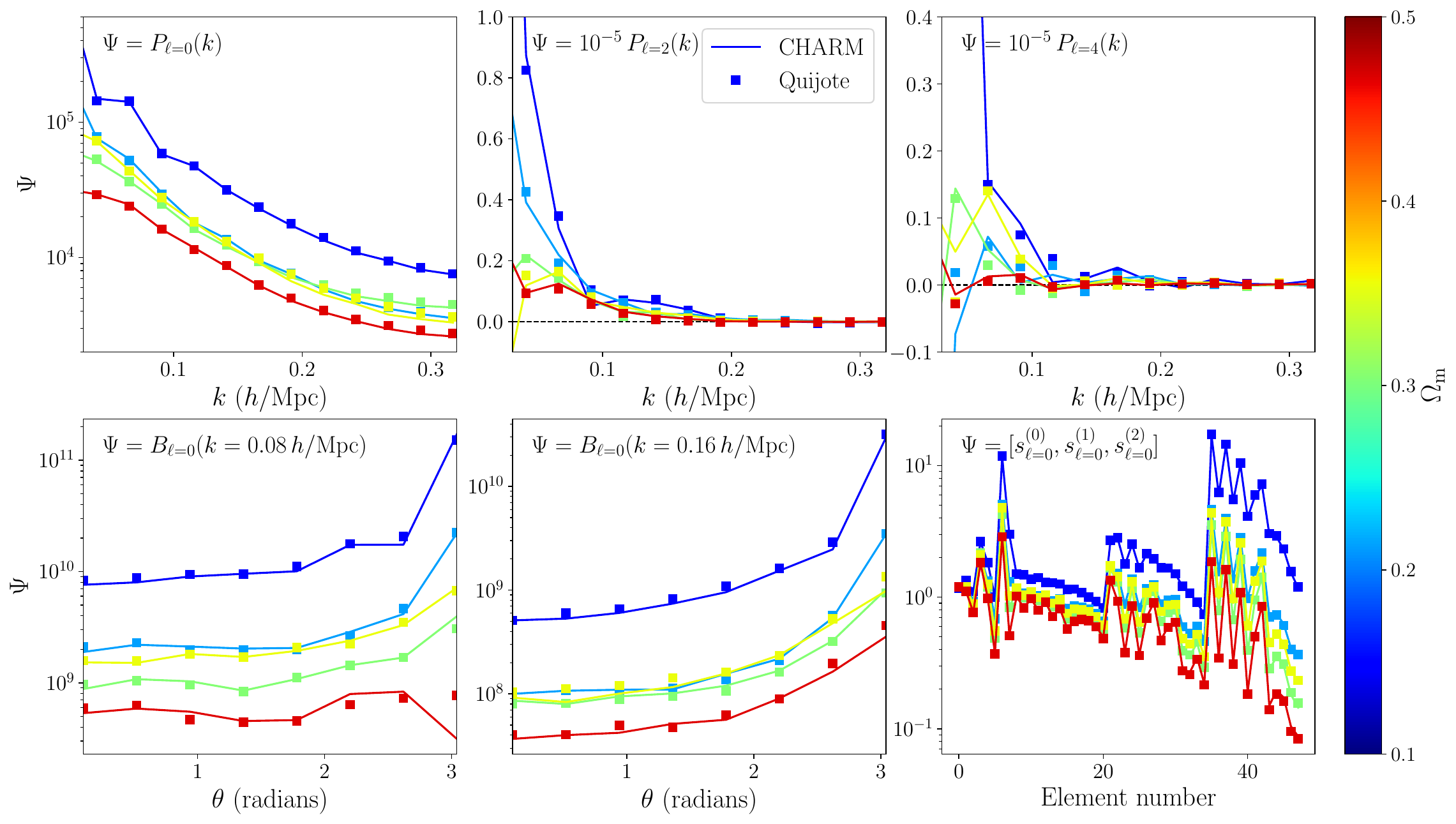}  
  \caption{\textbf{Global performance of the model with galaxies in redshift space:} Same as Fig.~\ref{fig:npt_rsd_halos} but for galaxies in redshift space and each curve for different cosmology uses a random set of HOD parameters.} 
  \label{fig:npt_rsd_galaxies}
\end{figure*}

\begin{figure*}
  \centering
 \includegraphics[width=1.\textwidth]{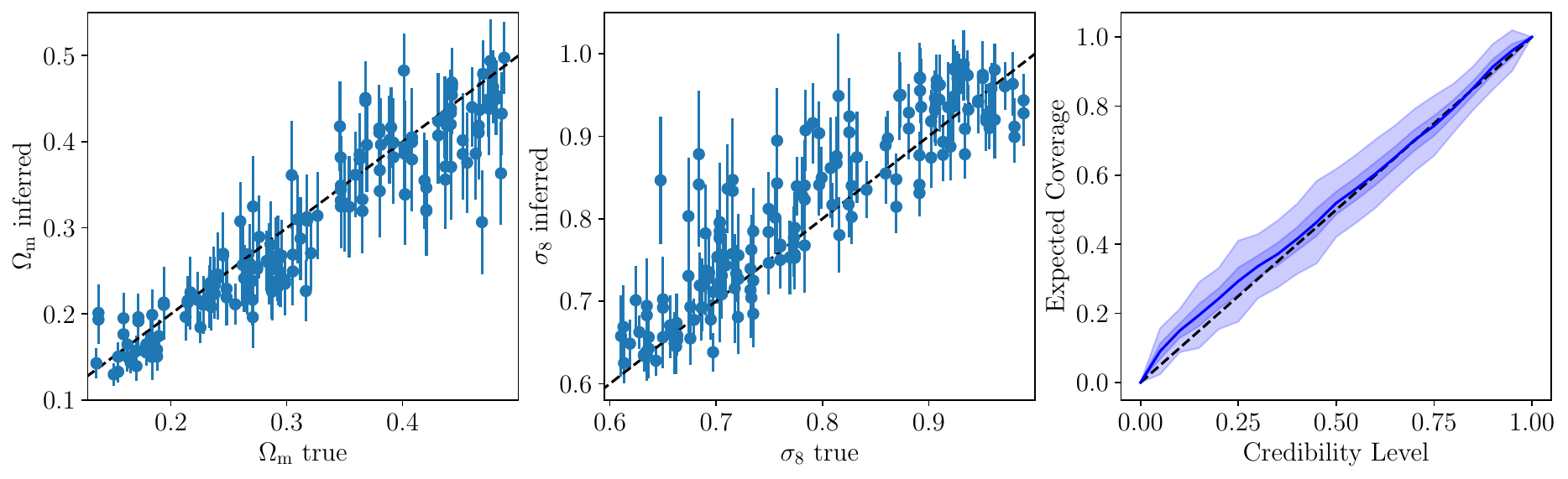} \caption{\textbf{Parameter inference with galaxies in redshift space:} Comparison of the true values of cosmological parameters in test simulations with the predicted values and their $1\sigma$ uncertainties using simulation-based inference with the redshift-space power spectra of galaxies ($P_{\ell = 0}, P_{\ell = 2}, P_{\ell = 4}$) for $k < 0.32 \, h/\mathrm{Mpc}$, monopole equilateral bispectra (with $\theta_{\rm k} \in [0.1, 3.04]$ radians) at $k \in \{0.06, 0.12, 0.32\} h/\mathrm{Mpc}$, and monopole first, second, and third-order wavelets: $s^{(0)}_{\ell = 0}, s^{(1)}_{\ell = 0}, s^{(2)}_{\ell = 0}$, as described in \S\ref{sec:stats_methods}. In the right panel, we also show the coverage of the posterior and its $1\sigma$ and $3\sigma$ uncertainty bands, using the methodology presented in \protect\cite{Lemos:2023:PMLR:}, with the diagonal line representing a well-calibrated posterior.}
  \label{fig:posterior}
\end{figure*}

\begin{figure*}
  \centering
 \includegraphics[width=1.\textwidth]{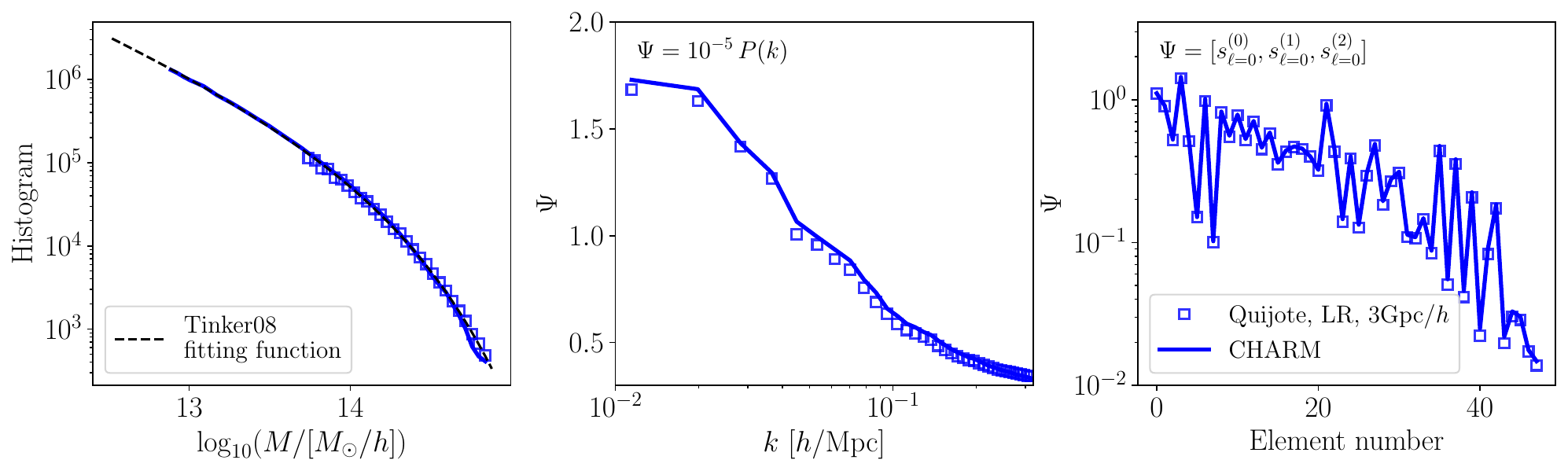}  
  \caption{Comparison of statistics of mock catalogs from \texttt{CHARM} with a low-resolution 3Gpc/$h$ box-size $N$-body simulation. In the left panel we compare the histogram of halo masses from the $N$-body simulation (unfilled square markers) which has a halo mass resolution limit of $\log (M/[ M_{\odot}/h ]) > 13.7$ and the one predicted from trained \texttt{CHARM} network (solid line) which has a mass limit of $\log (M/ [M_{\odot}/h ]) > 12.7$. We also compare them to the fitting function presented in \protect\cite{Tinker_2008} finding that \texttt{CHARM} can extrapolate to lower halo masses well. In the middle and right panels, we compare the power spectra and wavelet scattering transforms of halos in real space in the mass range resolved by the $N$-body simulation ($13.7 < \log (M/[M_{\odot}/h]) < 14.9$). }
  \label{fig:larger_vol}
\end{figure*}

\section{Results}
\label{sec:results}

The model is now trained using 1800 LH simulations and one fiducial-cosmology simulation from the Quijote suite (see \S~\ref{sec:data}).
We train the generative model for halo masses and numbers on four \texttt{Nvidia-H100} GPUs in 12 hours while training the model for halo velocities uses four \texttt{Nvidia-H100} GPUs for 8 hours. However, once the network is trained, obtaining the samples of halos and their velocities on test simulation takes less than a minute on a single \texttt{Nvidia-H100} GPU. We reserve the remaining 200 LH simulations as a test suite to compare the performance of the trained model. 

We test the performance of the model across three axes -- accuracy, inference, and scaling. We use a set of different statistical summaries described in \S~\ref{sec:stats_methods} to test the accuracy of the model, then use these summaries for galaxies in redshift space to perform inference on cosmological parameters and finally, we test how well the model scales to larger volume simulations. 

\subsection{Statistical summaries comparison}

Firstly, to test the performance of our trained network, we calculate the one-point statistics (histogram), two-point statistics (power spectrum), three-point statistics (bispectrum) as well as higher-order statistics (wavelet scattering) from \texttt{CHARM} and true $N$-body halo catalogs in test simulations with varying cosmologies. 

In Fig.~\ref{fig:1pt}, we compare the one-point statistics of the total number of halos ($N_{\rm halo}$), the mass of the heaviest halo ($M_1$), and the mass of the third heaviest halo ($M_3$). We compare these histograms in both low-density ($0 < \delta_{\rm m} < 2$) and high-density ($\delta_{\rm m} > 2$) environments in five test simulations with different cosmologies to highlight the large differences in these histograms and their dependence on the underlying dark matter density field. We observe a good match for all three histograms in both environmental conditions and all cosmologies. Note that this one-point comparison individually tests the performance of each of the three parts of the pipeline to infer the halo numbers and masses as mentioned in \S~\ref{sec:methods}. We see that, for all the cases where the number of halos is significant, our network is accurate at the percent level. 

To calculate beyond one-point statistics, we take the mock and truth halo catalogs in the 1000 Mpc/$h$ box and calculate the 3D weighted-density fields by applying a halo mass-dependent weight $w = (M/M')^{\alpha}$, where $M'=10^{14}M_{\odot}/h$ and $\alpha=0.7$. This power-law scaling mimics the weighting applied by the halo occupation distribution of SDSS-CMASS galaxies at $z=0.5$ \citep{reid2014, simbig_mock} and makes the statistics sensitive to environment-dependent halo count and halo masses. In Fig.~\ref{fig:npt_real_halos}, we compare the mock and true mean power spectrum, bispectrum, and wavelet coefficients from the test simulations in the real comoving space. We show the comparison for five of the 200 LH test simulations, which highlights the large variation in the statistics due to changing cosmologies. 
We find good agreement in all cases, where the performance of \texttt{CHARM} is either better than 5\% or within the cosmic variance uncertainty due to varying initial conditions. 

We now test the performance of the model when including the inference of the halo velocities as well. For this, we compare the power spectra multipoles, bispectra monopoles, and wavelet scattering monopoles in redshift space. We show the comparison of truth and mock catalog statistics using markers and solid lines respectively in Fig.~\ref{fig:npt_rsd_halos} for five cosmologies in the test simulation suite. We again find similar performance as for comparison in the real space. Note that, as the weighted power spectra, bispectra, and wavelet scattering transform probe the correlations between voxels -- as well as between halo counts, masses, and velocities -- it is a stringent test of the joint performance of all four stages of the network.
In Section~\ref{app:halo_stat_all}, we show the performance of the model using residuals of the summary statistics between the mock and true halo catalogs for all the 200 test simulations. 

Now we paste in galaxies on the halo catalogs from both the mock \texttt{CHARM} catalogs and true $N$-body simulations using a HOD formulation and measure the summary statistics in redshift space. Note that we additionally apply a Gaussian smoothing to the galaxy positions from both the mock and true catalogs (with a standard deviation of $\sigma_{\rm G} \sim 8\, {\rm Mpc}/h$) to account for the inability of our model to resolve halo positions below 8 Mpc/$h$ (see \S~\ref{sec:galaxies} for details). We compare different summary statistics for five test simulations (with varying cosmologies and random HOD parameters) in Fig.~\ref{fig:npt_rsd_galaxies}. We see that similar to the results for halo catalogs, we find that the model can infer the statistics of galaxies correctly up to the scales of $k < 0.32 \, h/{\rm Mpc}$.

\subsection{Cosmological inference}
We now use the summary statistics of 18000 mock galaxy catalogs obtained from \texttt{CHARM} (1800 LH cosmologies and 10 realizations with different values of the HOD parameters for each LH cosmology) which forms our training suite. Note that these are the same simulations that were used to train the \texttt{CHARM} network and we concatenate all the summary statistics described in \S~\ref{sec:methods} ($P_{\ell = 0}, P_{\ell = 2}, P_{\ell = 4}$, $B_{\ell=0}$ for $k\in [0.08, 0.16, 0.32]$, and $s^{(0)}_{\ell = 0}, s^{(1)}_{\ell = 0}, s^{(2)}_{\ell = 0}$) to form the input to the inference network. Moreover, as described in \S~\ref{sec:inference}, we train a set of 256 NPE and embedding networks with random hyperparameters to predict the five cosmological and five HOD parameters given the summary statistics and choose the five best networks with the lowest validation error. We form an ensemble of these five networks to jointly infer the parameter constraints based on the summary statistics measured directly from 200 test Quijote simulations (corresponding to the remaining 200 LH simulations and using a random HOD for each simulation).

In Fig.~\ref{fig:posterior}, we show the scatter between the true and predicted values of two of the cosmological parameters we expect to constrain with galaxy clustering statistics; $\Omega_{\rm m}$ and $\sigma_8$. We show the mean and 1$\sigma$ uncertainty on the parameters using the samples obtained from the learned posterior. We find that the network can accurately learn and constrain $\Omega_{\rm m}$ and $\sigma_8$ from the mock galaxy catalogs. Note that this inference marginalizes over other cosmological and HOD parameters. We also quantify the calibration of the uncertainty of our model in the parameter space by calculating the credibility level of the inferred parameters and comparing it against expected coverage. We use the method described in \cite{Lemos:2023:PMLR:}, which uses random points in the parameter space to estimate the expected coverage probabilities and also outputs the error in the estimated values using the bootstrap method. We show the coverage plot in the right panel of Fig.~\ref{fig:posterior} where we find that our inferred constraints agree with the diagonal line which corresponds to correctly calibrated posteriors.

\subsection{Applications to larger volume simulations}
The main advantage of a forward model like \texttt{CHARM} is that once trained, it can be applied to significantly larger volume simulations without the need for re-training since halo formation is only affected by the dark matter environment at scales less than 100\,Mpc/$h$ scales which we capture accurately here. To test this, we run a large volume $N$-body simulations with a box with 3 Gpc/$h$ side length. Due to computational constraints, we ran it at a lower resolution (with $1536^3$ particles) than the runs used to train \texttt{CHARM}. The $N$-body simulation takes approximately 8000 CPU hours and can only resolve halos with masses above $\log_{10}(M/[M_{\odot}/h]) > 13.7$. We run a paired \texttt{\texttt{FastPM}} simulation with the same initial condition in approximately 150 CPU hours. Afterward, we infer the halo catalogs from the trained \texttt{CHARM} network as described above which takes less than 2 GPU-minutes on one $\texttt{Nvidia-H100}$ GPU. We compare the statistics of the inferred and true halo catalogs in Fig.~\ref{fig:larger_vol}, showing the histogram of halo masses on the left, real space power spectra in the middle, and wavelet scattering transform on the right. The power spectra and wavelet scattering transform are calculated using halos resolved by the full $N$-body simulation, $13.7 < \log (M/[M_{\odot}/h]) < 14.9$. We see that the mock catalogs from \texttt{CHARM} match the statistics of halos in these larger volume simulations to within 5\% without any extra training. Moreover, the mock catalogs extend down to lower halo masses than the ones resolved in $N$-body simulation which we validate by comparing with the halo mass fitting function described in \cite{Tinker_2008} which uses several high-resolution simulations that can resolve low mass halos. Note that running a 3 Gpc/$h$ box-size $N$-body simulation that can resolve halos with masses of $\log_{10}(M/[M_{\odot}/h]) > 12.7$ would have taken approximately 80000 CPU-hours.

\section{Discussion}
\label{sec:discussion}
In this work, we describe how accelerating cosmological simulations and identifying collapsed dark matter structures (halos) are crucial for maximizing the information gained from current and upcoming galaxy surveys. We develop a multi-stage generative model that learns the relationship between the halo distribution and the surrounding dark matter distribution on large scales. Our approach uses a faster alternative to $N$-body simulations, which accurately captures dark matter densities on these scales. We validate our mock catalogs, obtained using \texttt{CHARM}, by comparing various statistics under different environmental conditions and cosmologies, demonstrating the method's accuracy.

We then paint galaxies onto the mock halo catalogs generated by the trained \texttt{CHARM} network and use the velocity information to calculate observable galaxy clustering statistics in redshift space. Comparisons with results from full $N$-body simulations show that, for scales of interest ($k < 0.32 h/$Mpc), \texttt{CHARM} produces accurate galaxy catalogs. Subsequently, we train a neural density network to infer cosmological and HOD parameters from observed galaxy clustering statistics and find that the calibrated posterior constrain cosmology accurately. Finally, we illustrate how \texttt{CHARM}, trained on high-resolution, small-volume $N$-body simulations can generate highly accurate halo catalogs in large-volume simulations at a fraction of the computational cost of running the full $N$-body simulation.

In this study, we trained the network with a physical resolution of 8 Mpc/$h$ and used a simple model of the galaxy-halo connection. Future work will focus on training the network on higher-resolution simulations to utilize even more information from galaxy catalogs. Additionally, incorporating more realistic halo-galaxy connection models will require information about secondary halo properties beyond mass, such as concentration and merger history \citep{jespersen2022}. We also trained the network at a fixed redshift of $z=0.5$, relevant for SDSS-CMASS-like galaxies. As we probe fainter galaxies with upcoming surveys, we will need to develop models that generalize to different redshifts. Moreover, in this study we have trained the model to recover the \texttt{Rockstar} halo catalogs with spherical overdensity mass definition, but the simulation based inference of cosmological parameters can be sensitive to this choice \citep{Modi:2024:fwdsbi}. However, a similar network can be trained to generate catalogs corresponding to other halo finders, such as friends-of-friends. Finally, as the trained generative model is differentiable, it can be optimized to find the maximum-a-posteriori estimate of the mock halo catalog conditioned on the surrounding dark matter properties, which will have a high cross-correlation coefficient with the true halo catalogs from the full $N$-body simulation. We plan to explore these developments in future studies.

\section*{Acknowledgments}
We thank Francois Lanusse for valuable discussions. SP thanks the CCA and CCM at the Flatiron Institute for hospitality while (a portion of) this research was carried out. The computations reported in this paper were performed using resources made available by the Flatiron Institute. The Flatiron Institute is supported by the Simons Foundation. 
This work is supported by the Simons Collaboration on ``Learning the Universe''. 

\section*{Data Availability}

Quijote data is publicly available at \href{https://quijote-simulations.readthedocs.io/en/latest/}{here}. The code to train the network, make all the predictions of summary statistics, and perform the cosmological inference is available \href{https://github.com/shivampcosmo/CHARM}{here}.
Direct access to already computed summary statistics and any other data used in this work can be requested by reaching out to any of the authors. 



\bibliographystyle{mnras}
\bibliography{fwm} 

\appendix
\label{app:halo_stat_all}
\begin{figure*}
  \centering
 \includegraphics[width=1.\textwidth]{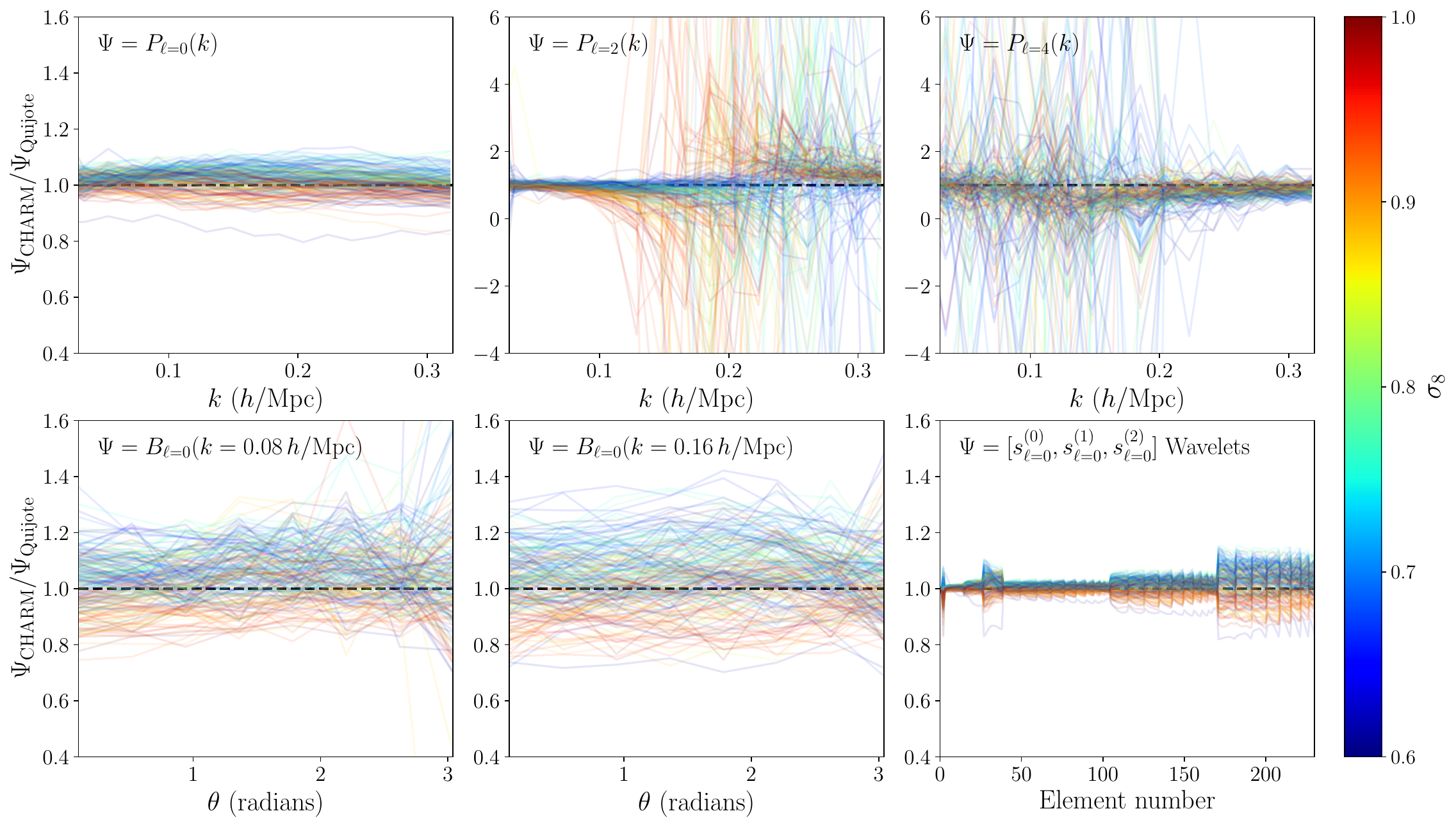}  
  \caption{\textbf{Residuals of the global performance of the model with halos in redshift space:} Same as Fig.~\ref{fig:npt_rsd_halos} but showing the residuals of each of the summary statistics. Here we show the residuals for all the 200 test simulations and color the lines with the cosmological parameter $\sigma_8$.}
  \label{fig:npt_rsd_halos_residual}
\end{figure*}
\section{Halo statistics residual performance}
In Fig.~\ref{fig:npt_rsd_halos_residual}, we compare the performance of the model by plotting the residual of the predicted and true summary statistics of halos in redshift space. We show the residuals for all 200 test LH Quijote simulations. We color each residual by the value of $\sigma_8$ parameter in that LH simulation. We find that the trained model works well either at approximately $5\%$ level or within the expected cosmic variance for a majority of the test simulations. We find a slight degradation of the performance for simulations having extreme values of the cosmological parameters but this does not bias our cosmological inference (see \S~\ref{sec:results}). 
It is important to note that the quadrupole ($P_{\ell = 2}(k)$) power spectra have a zero-crossing at $k \sim 0.2 \, h/{\rm Mpc}$ and hexadecapole ($P_{\ell = 4}(k)$) also crosses zero due to large noise, which makes the residual statistics to have large values.


\bsp	
\label{lastpage}
\end{document}